\documentclass[showpacs,twocolumn,prb]{revtex4-1}

\usepackage{graphicx}
\usepackage{amssymb}
\usepackage{amsmath}
\usepackage{color}
\usepackage{soul}
\usepackage{psfrag}
\usepackage{epsfig}
\usepackage{bbm}
\usepackage{bm}
\usepackage{ifthen}
\usepackage{calc}
\usepackage[latin1]{inputenc}
\usepackage[T1]{fontenc}
\usepackage{amsfonts}
\usepackage{amstext}
\usepackage{pst-all}
\usepackage[english]{babel}
\usepackage{subfigure}
\usepackage{rotating}
\usepackage{textcomp}
\usepackage{hyperref}
\definecolor{darkblue}{rgb}{0.0,0.0,0.3}
\hypersetup{colorlinks,breaklinks,linkcolor=darkblue,urlcolor=darkblue,anchorcolor=darkblue,citecolor=darkblue}

\newcommand{\bld}[1]{\boldsymbol #1}
\newcommand{\ket}[1]{| #1 \rangle}

\newcommand{\ew}[1]{\langle #1 \rangle}
\newcommand{\beq}{\begin{eqnarray}}
\newcommand{\eeq}{\end{eqnarray}}

\newcommand{\op}[2]{| #1 \rangle \langle #2 |}

\newcommand{\eq}[1]{Eq.~(\ref{#1})}
\newcommand{\fig}[1]{Fig.~\ref{#1}}

\begin{document}

\title{Spin entangled two-particle dark state in quantum transport through coupled quantum dots}
\author{Christina P\"oltl}
\email[Current address: School of Physical and Chemical Sciences, Victoria University of Wellington, P.O. Box 600, Wellington 6140, New Zealand, E-mail: ]{ christina.poeltl@vuw.ac.nz}
\affiliation{
  Institut f\"ur Theoretische Physik,
  Hardenbergstr. 36,
  TU Berlin,
  D-10623 Berlin,
  Germany
}
\author{Clive Emary}
\affiliation{
  Institut f\"ur Theoretische Physik,
  Hardenbergstr. 36,
  TU Berlin,
  D-10623 Berlin,
  Germany
}
\author{Tobias Brandes}
\affiliation{
  Institut f\"ur Theoretische Physik,
  Hardenbergstr. 36,
  TU Berlin,
  D-10623 Berlin,
  Germany
}

\date{\today}

\begin{abstract}
We present a transport setup of coupled quantum dots that enables the creation of spatially separated spin-entangled two-electron dark states. We prove the existence of an entangled transport dark state by investigating the system Hamiltonian without coupling to the electronic reservoirs.  
In the transport regime, the entangled dark state, which corresponds to a singlet, has a strongly enhanced Fano factor compared to the dark state, which corresponds to a mixture of the triplet states. 
Furthermore, we calculate the concurrence of the occupying electrons to show the degree of entanglement in the transport regime.   
\end{abstract}
\maketitle

\section{Introduction}
 The investigation of dark states (DSs) has a long-standing tradition in quantum optics both experimentally \cite{O-DS1} and theoretically.\cite{O-DS3,O-DS5} 
In recent years, there have been numerous approaches to translate this quantum optical phenomenon into electronic transport. \cite{BrRe00-DS1,Bra05-DS2,FSF03-DS3,MEB06-DS-TQD1,GMB06-DS-TQD2,Ema07-DS-TQD3,BSP10-TQD-DS,Busl2010830,DGK10-DS-Bun,BP12-DS-TQD}  
Here, the term DS is used when the current-carrying particles, in general, electrons, are trapped in a coherent superposition of states that is decoupled from the collector. The particle flow through the system is blocked, as no further electrons can enter the system due to the Coulomb blockade (CB). 
The first concepts used similar system setups as in quantum optics and included interactions with microwaves in order to create the {DS}. \cite{BrRe00-DS1,Bra05-DS2}    

A triple quantum dot (TQD) with a single excess electron was the first system  where an {\it all-electronic} DS was found by Michaelis {\it et al.} \cite{MEB06-DS-TQD1,GMB06-DS-TQD2} hence, the system is driven into the DS purely due to the coupling to the electronic reservoirs. 
Michaelis {\it et al.} showed the coherent trapping effect in the {TQD} and its destabilization due to charge fluctuations. This electronic {DS} was found to give rise to an enhanced Fano factor \cite{GMB06-DS-TQD2} above the Poissonian value $F>1$.  
The influence of a magnetic field on the {DS} formation in the {TQD} was studied in Ref.~[\onlinecite{Ema07-DS-TQD3,Busl2010830}], and Weymann {\it et al.}\cite{WBB11-TQD-DS-co} have presented the effects of co-tunneling on the DS formation.
The influence of phonon interaction on the  dark state formation in the {TQD} was studied in Ref.~[\onlinecite{DSP11-DS-nm}], and
Ref.~[\onlinecite{LOL11-DS-nano-res}]  showed how the  {TQD} dark state can be used as a nanomechanical resonator cooler.  

We have previously shown\cite{PEB09} that transport {DS} are not solely an issue of strong Coulomb blockade systems with only a single excess electron. In a {TQD} with a second excess electron, a two-electron {DS} can be found for certain configurations. 
This two-electron {DS} can also be used as a nanomechanical resonator cooler.\cite{ZLF-TQD-2el-DS} 
That electronic {DS}s also occur in interaction with other blockade phenomena is shown in Ref.~[\onlinecite{BSP10-TQD-DS}] where a mixture of a spin blockade and a single-electron {DS} was shown to lead to a quasi-two-electron {DS}.

While the two-electron DS in the single {TQD} of Ref.~[\onlinecite{PEB09}] is a product state of two single-electron {DS}s, in this paper we introduce a system that enables the preparation of a spin-entangled two-electron {DS}. 
For this aim, we consider two triple quantum dots with a single excess electron in each dot.  
A possible application of this setup is the creation of spacially separated entangled electrons on demand.

The structure of this paper is the following:   
After introducing the model in Sec. \ref{sec:model}, we investigate the existence of dark states in the closed system without coupling to the electronic reservoirs in Sec. \ref{sec:closed}. The transport properties, namely, stationary current and Fano factor, are discussed in Sec. \ref{sec:trans}, and in order to show the degree of entanglement in the transport regime, we calculate the concurrence in Sec.  \ref{sec:concur}.

\section{Model \label{sec:model}}
\begin{figure*}[t!]
 \centering
\includegraphics[width=0.95\textwidth]{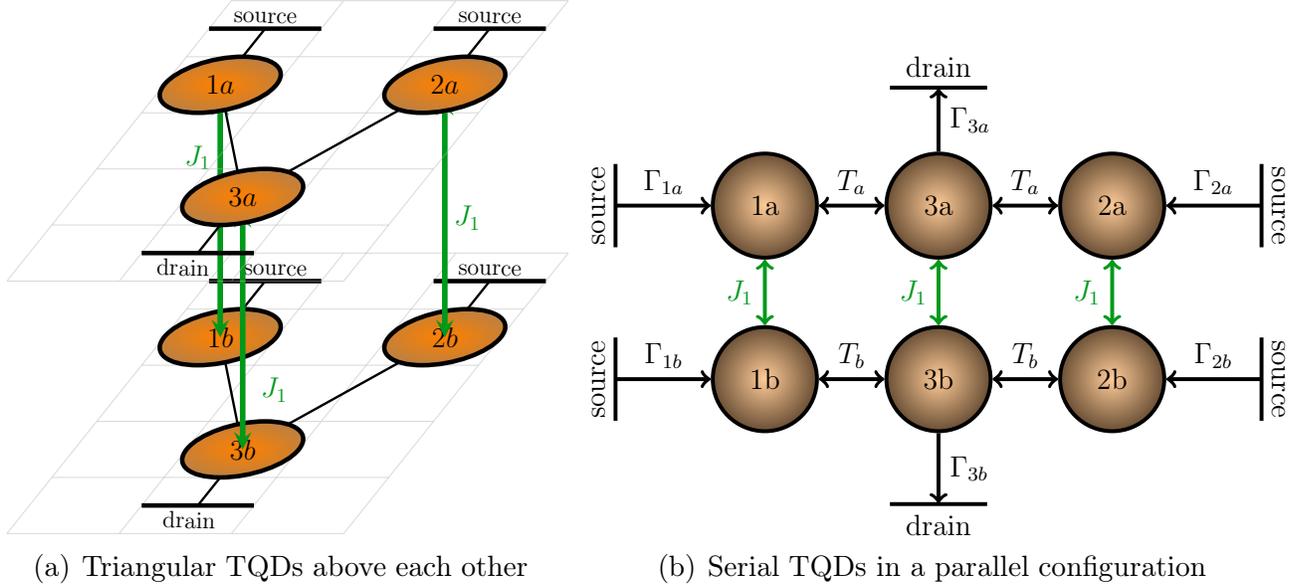}
\caption[Two {TQD}s above each other]{Both {TQD}s are in the strong Coulomb blockade regime such that up to one electron is allowed in each {TQD}. The electrons in the TQDs interact with each other capacitively due to the charging energies $U_{ij}$ and due to the exchange interaction switching the spins of the electrons. (The exchange interaction $J_1$ is indicated  in the sketches.)  Each {TQD} is connected to two sources and one drain. Here, two possible configurations are shown where (a) the two {TQD}s are triangular and lie above each other and (b) the TQDs are serial and parallel to each other. 
}
\label{fig:2TQD}
\end{figure*}

Fig.~\ref{fig:2TQD} shows two possible configurations of the two TQDs.
Both {TQD}s are in the strong Coulomb blockade regime such that, up to one electron is allowed in each {TQD}. The {TQD}s are close together, therefore we have a finite charging energy between the {TQD}s. Furthermore, we introduce an isotropic exchange interaction acting between the two {TQD}s.
The complete closed system Hamiltonian $\hat{H}_D$ is given by 
\begin{eqnarray}
 \hat{H}_D=\hat{H}_{\text{TQD},a}+\hat{H}_{\text{TQD},b}+\hat{U}+\hat{J}, \label{H-syst}
\end{eqnarray}
where the {TQD} Hamiltonians are given by
\begin{align}
\hat{H}_{\text{TQD,}A} =& \sum_{i=1}^3 \sum_{\sigma}E_{i,A}\hat{n}_{i \sigma,A} 
+T_{A}\sum_{j=1}^2 \sum_{\sigma}(d^{\dagger}_{j \sigma,A}d_{3 \sigma,A}+h.c),
\end{align}
with $A\in \{a,b\}$,  $d_{i\sigma,A}$ is the annihilation operator, and $\hat{n}_{i\sigma,A}$ is the corresponding number operator of an electron in quantum dot (QD) $i$ of {TQD} $A$ with spin $\sigma$. We assume spin-independent energy levels and denote the energy of the single-electron level of a quantum dot {QD}$i$,$A$ as $E_{i,A}$.
In the following we set all $E_{i,A}=0$.
The levels in {QD}$1,A$ and {QD}$2,A$ are coupled coherently to {QD}$3,A$ with a tunnel amplitude $T_A$.
The charging energy is the capacitive part of the Coulomb interaction 
and given by 
\begin{align}
\hat{U} = \sum_{i,j
}^3 \sum_{\sigma, \sigma'} U_{i j } \hat{n}_{i \sigma,a} \hat{n}_{j \sigma',b} \, , 
\label{DS:ent-TQD_coul} 
\end{align}
where $U_{i j }$ is the additional charging energy needed to add an electron to QD $i$ of TQD $a$ when QD $j$ of TQD $b$ is occupied with one electron.
In this setup, an electron in TQD $a$ always interacts with an electron in TQD $b$, as both TQDs are in the strong Coulomb blockade. 
Therefore, terms for having two electrons in a single TQD are not included in $\hat{U}$, as they are assumed to be far above the transport window and not relevant for the transport.

The isotropic exchange energy is  
\begin{align}
 \hat{J} &=\sum_{i,j} J_{ij}(\bld{\sigma}_{i,a} \cdot \bld{\sigma}_{j,b})\mathord
=\sum_{i,j} J_{ij}(\sigma_{i,a}^x  \sigma_{j,b}^x \mathord+\sigma_{i,a}^y  \sigma_{j,b}^y \mathord+\sigma_{i,a}^z \sigma_{j,b}^z) \nonumber \\
&=\sum_{i,j} J_{ij} \big(\sigma_{i,a}^z \sigma_{j,b}^z +2(\sigma_{i,a}^+  \sigma_{j,b}^- +\sigma_{i,a}^-  \sigma_{j,b}^+ )\big) \nonumber \\
&=\sum_{i,j} J_{ij} \big[(\hat{n}_{i \uparrow,a}-\hat{n}_{i \downarrow,a})(\hat{n}_{i \uparrow,b}-\hat{n}_{i \downarrow,b})
\nonumber \\ &
+
2(d^{\dagger}_{i \uparrow,a}d_{i \downarrow,a} d^{\dagger}_{j \downarrow,b}d_{j \uparrow,b}+
d^{\dagger}_{i \downarrow,a}d_{i \uparrow,a} d^{\dagger}_{j \uparrow,b}d_{j \downarrow,b} ) \big], \label{exc-int}
\end{align}
where $\bld{\sigma}$ are the Pauli-matrices, $i$ and $j$ label the {QD}s of the {TQD} $a$ or $b$, and $J_{ij}$ are the exchange constants. In the following, we set $J_{ii}=J_1$ and  $J_{ij}=J_2$, $i\neq j$.  In this paper, we treat this exchange interaction as a part of the Coulomb interaction between the electrons.\cite{foot1}                    
The system Hamiltonian in the localized basis can be found in Appendix \ref{App:2TQD-DS}. Later 
we will refer to the two-electron Hamiltonian blocks $H_{\sigma a \sigma' b}$ for the different spin configuration 
and the exchange interaction blocks $\bar{J}$ defined in this appendix. 

Each {TQD} is connected to three electron reservoirs that are described with the Hamiltonian
\begin{eqnarray}
\hat{H}_{\text{res}} &=& \sum_{\alpha,A} \sum_{k,\sigma} \epsilon_{\alpha k,A} c^\dagger_{\alpha k\sigma,A} c_{\alpha k \sigma,A}, \label{DS:ent-res}
\end{eqnarray}
where  $\alpha := \{1,2,3 \}$ labels the reservoirs ($1,2 =$ source, $3=$ drain) and $c^\dagger_{\alpha k\sigma,A}$ is the creation operator of an electron with spin $\sigma$ in mode $k$ of reservoir $\alpha$ of TQD $A$. 
The {TQD} and the reservoirs are connected by the tunnel Hamiltonian 
\begin{align}
\hat{H}_{\text{T}} =& \sum_{\alpha,A} \sum_{k,\sigma}  V_{\alpha k,A}c^{\dagger}_{\alpha k\sigma,A}d_{\alpha \sigma,A} +\mathrm{h.c.}
\label{DS:ent-coup}
.
\end{align}
We assume spin-independent reservoir energies $\epsilon_{ik,A}$ and tunneling amplitudes $V_{ik,A}$.

\section{Closed system \label{sec:closed}}
We consider a transport system operating in the high-bias regime.\cite{GuPr96-hb-ma,Gur98-hb-ma} This regime can be assumed when all relevant energy levels of the system lie well within the transport window and the temperature of the transport device is low. In the high-bias regime, the transport is  unidirectional, hence all electrons enter the system from the source leads and leave the system through the drain leads.  
In such a system, the formation of a transport DS $ |\varPsi_{D}\rangle $ is only possible if the system Hamiltonian $\hat{H}_S$ fulfills certain conditions. 
But the formation of the DS can be destroyed due to decoherence\cite{BrRe00-DS1,SM12-no-DS} or avoided for special coherent system-bath couplings,\cite{SM12-no-DS} even when these conditions are fulfilled.   
A transport DS can be found when the system Hamiltonian-block, with the most excess electrons has an eigenstate $ |\varPsi_{D}\rangle $  without finite occupation on the QD(s) which is (are) coupled to the collector(s). For the two-TQD setup, this means that we search for an eigenstate  in the two-electron sector without occupation on {QD}3 of both TQDs,
\begin{align}
 \ew{ \varPsi_{D}|\hat{n}_{3 \sigma,A}|\varPsi_{D}}=0, \quad \forall A, \sigma.
\end{align}
In the transport regime, such an eigenstate leads in general to a current blockade, where the stationary current of the system drops to zero when the DS becomes occupied. In the following discussion we set the charging energy $U_{21}=U_{12}$.

\begin{figure*}[t]
  \begin{center}
\includegraphics[width=1\textwidth]{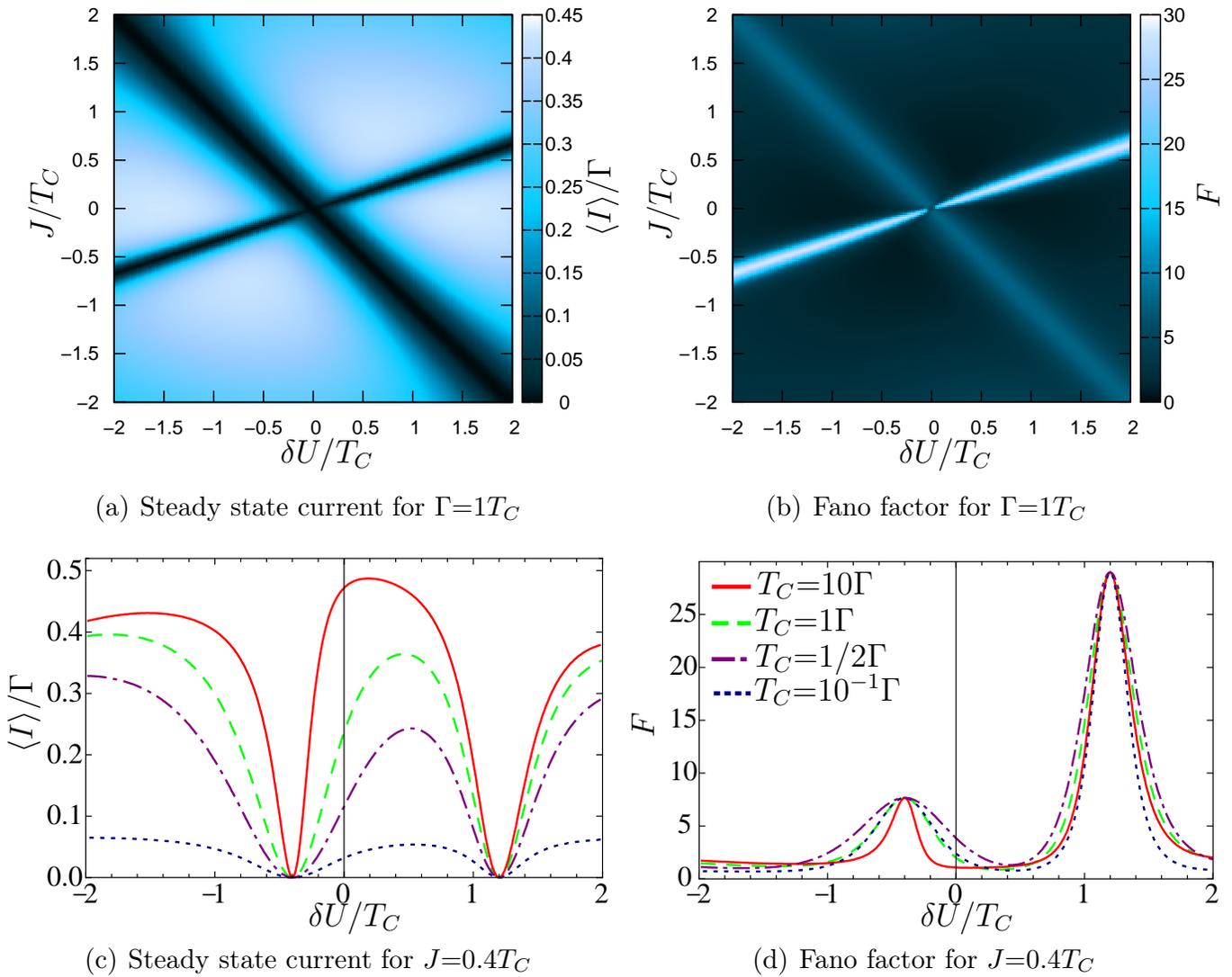}
  \end{center}  
\caption[Current and Fano factor of the two-{TQD} system with the spin-entangled two-electron {DS}s.]{(a) Steady-state current $\ew{I}/\Gamma$ as a function of the exchange energy $J$  and $\delta U$ normalized by $T_C$ for $\Gamma/T_C=1$. The thin dark line (zero-current line) in the density plot indicates the formation of the  singlet-DS and the thick dark line indicates the formation of the triplet-{DS}. (b) Fano factor (corresponding to (a)) is highly super-Poissonian around both {DS}s. But the Fano factor near the singlet-DS is around four times higher than around the triplet-DS. 
(c) Current as a function of $\delta U/T_C$, for different coupling strength $T_C/\Gamma$ at $J=0.4T_C$. The curve for $T_C=1\Gamma$ corresponds to a section through the density plot above.  (d) The Fano factors show that the maximum value of the Fano factor near both {DS} is independent of the ratio $T_C/\Gamma$. Parameter: $V=10T_C$.                      
     }\label{Fig:2-TQD-DS-CF}
\end{figure*}

\subsection{Closed system without exchange interaction}
We begin by looking at the system without exchange interaction and set $J_1=J_2=0$, such that the charging energy is the only influence that exists between the two {TQD}s. Without exchange energy, the two-electron sector of the system Hamiltonian consists of four  blocks which are uncoupled to each other. 
Each block corresponds to one of the four possible spin configurations of the occupying electrons. In order to find a transport DS, we are searching for eigenstates without occupation on the third dots in one of these four blocks,  hence eigenstates of the form
\begin{align}
 |\varPsi_{D}\rangle =& (
a_1 d^\dag_{1\sigma,a}d^\dag_{1\sigma',b}
+a_2 d^\dag_{1\sigma,a}d^\dag_{2\sigma',b}\nonumber \\ &
+a_3 d^\dag_{2\sigma,a}d^\dag_{1\sigma',b}
+a_4 d^\dag_{2\sigma,a}d^\dag_{2\sigma',b} )\ket{0},
 \label{Psi0J0}
\end{align} 
with $|a_1|^2+|a_2|^2+|a_3|^2+|a_4|^2=1$. But these blocks are $9\times9$ matrices and it is not possible in general,  to calculate all eigenstates analytically in order to prove that a dark state exists.  
However, the state $|\varPsi_{D}\rangle$ must fulfill, in the spin-$\sigma\sigma'$ sector of the localized basis, the condition
\begin{eqnarray}
 (H_{\sigma a \sigma' b}-\lambda_D \mathbbm{1}) |\varPsi_{D}\rangle = 0. \label{DS-condition-1}
\end{eqnarray} 
The explicit form of the $H_{\sigma a \sigma' b}$ can be found in Appendix \ref{App:2TQD-DS}.  
Each of these four blocks has equal entries because we assume spin-degenerate single-particle energies.  
Hence, a DS in one of the blocks is degenerated with the DSs at the same energy $\lambda_D$ in the other three blocks.\cite{foot-M}   
In the two-TQD setup without exchange interaction, we find a {DS} 
\begin{eqnarray}
 |\varPsi_{D,\sigma \sigma'}\rangle &=& 
\frac{1}{2} (d^\dag_{1\sigma,a}-d^\dag_{2\sigma,a}) (d^\dag_{1\sigma',b}-d^\dag_{2\sigma',b})\ket{0}, 
 \label{Psi0a}
\end{eqnarray} 
in each block, when the charging energies $U_{12}=U_{11}=U_{22}$ are equal to the {DS}-eigenenergy, $\lambda_D=U_{12}$. 
The DS of the transport system is then a mixture of the four degenerate states  of the closed system.

\subsection{Closed system with exchange interaction}
With a finite isotropic exchange interaction, the two-electron blocks with opposite spins couple to each other, while the blocks with equal spins remain uncoupled. Now a two-electron DS is found either when        
\begin{eqnarray}
 (H_{\sigma a \sigma b}+\bar{J}-\lambda_D \mathbbm{1}) |\varPsi_{D}\rangle = 0, \quad \text{or} \nonumber \\
 \left(\begin{pmatrix}
 H_{\uparrow a \downarrow b}-\bar{J} & 2\bar{J}      \\
2 \bar{J} & H_{\downarrow a \uparrow b }-\bar{J}  
\end{pmatrix}-\lambda_D \mathbbm{1}\right) |\varPsi_{\bar{D}}\rangle = 0,
\end{eqnarray}
with 
\begin{align}
 |\varPsi_{\bar{D}}\rangle =& (
a_1 d^\dag_{1\uparrow,a}d^\dag_{1\downarrow,b}
+a_2 d^\dag_{1\uparrow,a}d^\dag_{2\downarrow,b}
+a_3 d^\dag_{2\uparrow,a}d^\dag_{1\downarrow,b}
\nonumber \\ &
+a_4 d^\dag_{2\uparrow,a}d^\dag_{2\downarrow,b} 
+b_1 d^\dag_{1\downarrow,a}d^\dag_{1\uparrow,b}
+b_2 d^\dag_{1\downarrow,a}d^\dag_{2\uparrow,b}
\nonumber \\ &
+b_3 d^\dag_{2\downarrow,a}d^\dag_{1\uparrow,b}
+b_4 d^\dag_{2\downarrow,a}d^\dag_{2\uparrow,b}
)\ket{0}, 
 \label{Psi0}
\end{align} 
and $|a_1|^2+|a_2|^2+|a_3|^2+|a_4|^2+|b_1|^2+|b_2|^2+|b_3|^2+|b_4|^2=1$. DSs of the form of \eq{Psi0a} still exist when the two occupying electrons have equal spin, 
but now the charging energies have to fulfill the condition $U_{1 1} =  U_{2 2 } = J_2 -J_1 + U_{1 2 }$ and the eigenenergy is shifted to $\lambda_D=J_2+U_{1 2}$.

In the opposite spin sector, we find now two DSs with different conditions for the charging energies and with different eigenenergies.
For $\lambda_D=-3J_2 + U_{1 2}$ and $U_{11} =  U_{22} = 3J_1-3J_2  + U_{12}$, the {DS} is a singlet state,
\begin{eqnarray}
 |\varPsi_{D,-}\rangle &=& 
\frac{1}{\sqrt{2}} (\ket{\varPsi_{D,\uparrow \downarrow}}-\ket{\varPsi_{D,\downarrow \uparrow}}),
 \label{Psi0J1e}
\end{eqnarray} 
and 
for $\lambda_D=J_2+U_{1 2}$ and  $U_{1 1} =  U_{2 2 } = J_2 -J_1 + U_{1 2 }$, the {DS} is a  triplet state,
\begin{eqnarray}
 |\varPsi_{D,+}\rangle &=& 
\frac{1}{\sqrt{2}} (\ket{\varPsi_{D,\uparrow \downarrow}}+\ket{\varPsi_{D,\downarrow \uparrow}}).
 \label{Psi0J1f}
\end{eqnarray} 
These two dark states are entangled with respect to the spin of the electrons. 
In the transport regime, only the singlet-DS can be prepared as a pure state, since it is not degenerate with any other DSs.
The entangled triple-DS given by \eq{Psi0J1f} is degenerate 
 with the two DSs in the equal spin sectors, which also correspond to the other two states of the triplet. 
Not only the degeneracy of the eigenstate is broken, but also the conditions for charging and exchange energies are different for singlet-DS and triplet-DS. 
This enables one to prepare the pure transport singlet-DS in the high bias regime, where all relevant system states are well within the transport window.
If the degeneracy is broken, but the condition for the charging energies remains equal for all four DSs, only for certain finite transport window configurations it would be possible to prepare the singlet-DS as a pure state.   

\subsection{Asymmetric setups}
The high symmetry of our system Hamiltonian given by \eq{H-syst}  might not be realizable in realistic setups. We therefore like to address, the conditions under which the entangled-DS formation is still possible in asymmetric setups. 

Without a finite coupling strength $T_{12}$ between dot $1$ and dot $2$ of each TQD, the symmetry between the coupling of dot $1$ to dot $3$ and dot $2$ to dot $3$ is indeed necessary and independent of the additional $T_{12}$. However, with a finite coupling strength $T_{12}$, which exists especially in a realistic triangular setup, asymmetric combinations become possible. 
The exact relations between the coupling strengths of each TQD are completely analogous to the single-electron DS in a single TQD, as found in Ref. 
[\onlinecite{Ema07-DS-TQD3}]. The parameters can be chosen separately in each of the TQDs and the entanglement of the electrons is not affected, by asymmetric coupling parameters. 

Apart from the coupling strength, the exchange energies between the dots can also differ for all possible combinations of occupations. We concentrate here on the case where only the exchange interaction in the QDs above each other differ. These exchange interactions are then denoted as $J_{ii}$ according to \eq{exc-int}. The DS formation is completely independent of the value of $J_{33}$. If $J_{11}$ and $J_{22}$ differ, we find the singlet-DS when $U_{11}=U_{12}+3 J_{11}$, $U_{22}=U_{12}+3 J_{22}$ and $\lambda_D=U_{12}$. Once again the entanglement remains unaffected by the asymmetry of the parameters.

\section{Transport properties \label{sec:trans}}
The considered transport setup is such that all electrons enter the TQDs from the source leads with the rates $\Gamma_{iA}$, $i\in\{1,2\}$, and $A\in\{a,b\}$ depending on the QD and TQD in which the electrons tunnel and leave the TQDs by tunneling into the drain lead with the rate $\Gamma_{3A}$.  We assume that all considered energy levels of the two TQDs lie well within the transport window. We can therefore use a generalized master equation in Lindblad form  \cite{GuPr96-hb-ma,Gur98-hb-ma} to described the transport through the two-TQD setup,
\begin{align}
\dot{\rho}=-i[\hat{H}_D,\rho]+\sum_X\left(D_X\rho D^\dagger_X-\frac{1}{2}D^\dagger_XD_X\rho-\frac{1}{2}\rho D^\dagger_XD_X\right).  \label{TQD-ent:Trans_Lin}
\end{align}
The explicit form of the 12 coupling terms $X=jA \sigma $, with $j\in\{1,2,3\}$, $A\in\{a,b\}$, $\sigma\in\{\uparrow,\downarrow \}$, can be found in Appendix \ref{App:2TQD-DS-B}.
In order to calculate the stationary current and the second-order zero-frequency Fano factor
we rewrite \eq{TQD-ent:Trans_Lin} in Liouville space $\dot{\bld{\rho}}(\chi)=(\mathcal{W}_0+\mathcal{J}e^{i\chi})\bld{\rho}(\chi)$ and introduce a counting field.\cite{LeLe93-FCS,LLL96-FCS,BaNa03-FCS,MEB10-FCS} This counting field enables one to introduce the cumulant generating function of the current distribution
\begin{align}
 {\cal F}(\chi,t)= \ln\Big(\text{Tr}_{D}\big\{ e^{({\cal W}_0 + 
   {\cal J} e^{i\chi })(t-t_0)}\rho(t_0)\big\} \Big),
\end{align}
where $\text{Tr}_D\{\cdots\}$ corresponds to the trace of the density matrix. The $n$th-order zero-frequency current correlation \cite{MEB10-FCS} is then evaluated by 
\begin{align}
\ew{S^{(n)}}=\frac{\text{d}}{\text{d}t}\frac{\partial^n}{\partial (i\chi)^n} 
{\cal F}(\chi,t) |_{\chi=0,t\rightarrow\infty}. 
\end{align}
The second-order zero-frequency Fano factor is then defined as
\begin{align}
 F=\frac{\ew{S^{(2)}}}{\ew{S^{(1)}}},
\end{align}
the second-order current correlation functions, normalized by the stationary current ($\ew{S^{(1)}}=\ew{I}$). Since we are interested in the total current and noise through the system, we count the electrons tunneling from both {TQD}s. 
With this, the jump operator becomes $\mathcal{J}\bld{\rho}=\sum_{A,\sigma}D_{3A\sigma}\rho D^\dagger_{3A\sigma}$ and $\mathcal{W}_0$ correspond to the other terms of the Lindblad equation \eq{TQD-ent:Trans_Lin}. We could also count the electrons leaving each {TQD} separately. However, for the parameter setting for which we calculate the steady-state current and Fano factor in this paper, the results would be simply half of the total current and Fano factor. 

In the following discussion of steady-state current and Fano factor, we set $J_1=J$, $J_2=0$, $T_a=T_b=T_C$, $\Gamma_{1A} =\Gamma_{2A}=\Gamma_{3A}=\Gamma$, $U_{i i} =  U$, and $U_{i j } =  V$, for $i\neq j$.  We then introduce $\delta U=U-V$ as the difference between intra-charging and inter-charging energy. Figure \ref{Fig:2-TQD-DS-CF} shows the total steady-state current and Fano factor of the two-{TQD} setup. \fig{Fig:2-TQD-DS-CF}(a) is a density plot of the current $\ew{I}/\Gamma$ as a function of exchange interaction $J$ and charging energy difference $\delta U$ normalized by $T_C$. The formation of both dark states, singlet-DS as well as triplet-DS, can be seen as dark lines running through the density plot. However, the width of the anti-resonance in the current around the triplet-DS for this parameter setting is much broader than for the singlet-DS. At $\delta U =J=0$, where the current valleys cross each other, the singlet-DS and the triplet-DSs of the closed system as well as of the transport 
system live in a 
degenerated 
subspace. The transport DS at $\delta U =J=0$ is therefore a mixture of the three triplet-DSs and the singlet-DS.\cite{foot2}  
\fig{Fig:2-TQD-DS-CF}(b) shows the corresponding Fano factor to the current 
density plot. Although the Fano factor reaches near both dark states' highly super-Poissonian values,  the maximum value around the singlet-DS is strongly increased compared to the triplet-DS. 

\fig{Fig:2-TQD-DS-CF}(c) shows the current as a function of the charging energy difference $\delta U/\Gamma$ for different ratios of $T_C/\Gamma$. The current increases asymptotically  with increasing $T_C$. Hence, the not-shown current  and Fano factor curves for $T_C=100\Gamma$ almost coincide with current and Fano factor curves for $T_C=10\Gamma$. The width of the current valley around the {DS}s decreases for both {DS} with increasing $T_C$, but again remains finite for $T_C\rightarrow\infty$. Apart from that, the valley around the triplet-DS decreases more strongly than the valley around the singlet-DS.

In \fig{Fig:2-TQD-DS-CF}(d), the corresponding Fano factor to \fig{Fig:2-TQD-DS-CF}(c) is shown. The maximum value of the Fano factor near both {DS}s is independent of the ratio $T_C/\Gamma$ and highly super-Poissonian. But the value of the Fano factor around the singlet-DS is approximately four times higher than the value around the triplet-DS. Similar features are found by Burkard {\it et al.} in Ref.~[\onlinecite{BLS00-AB-B-Fw}].
The width of the Fano factor resonance is widest for $T_C=1/2 \Gamma$ and decreases for both smaller and higher values of $T_C/\Gamma$, as shown in the plot. Therefore, the width of the Fano factor resonance is not simply decreasing for higher values of $T_C/\Gamma$ as for the corresponding current.

\begin{figure}[t]
 \centering
\includegraphics[width=\columnwidth]{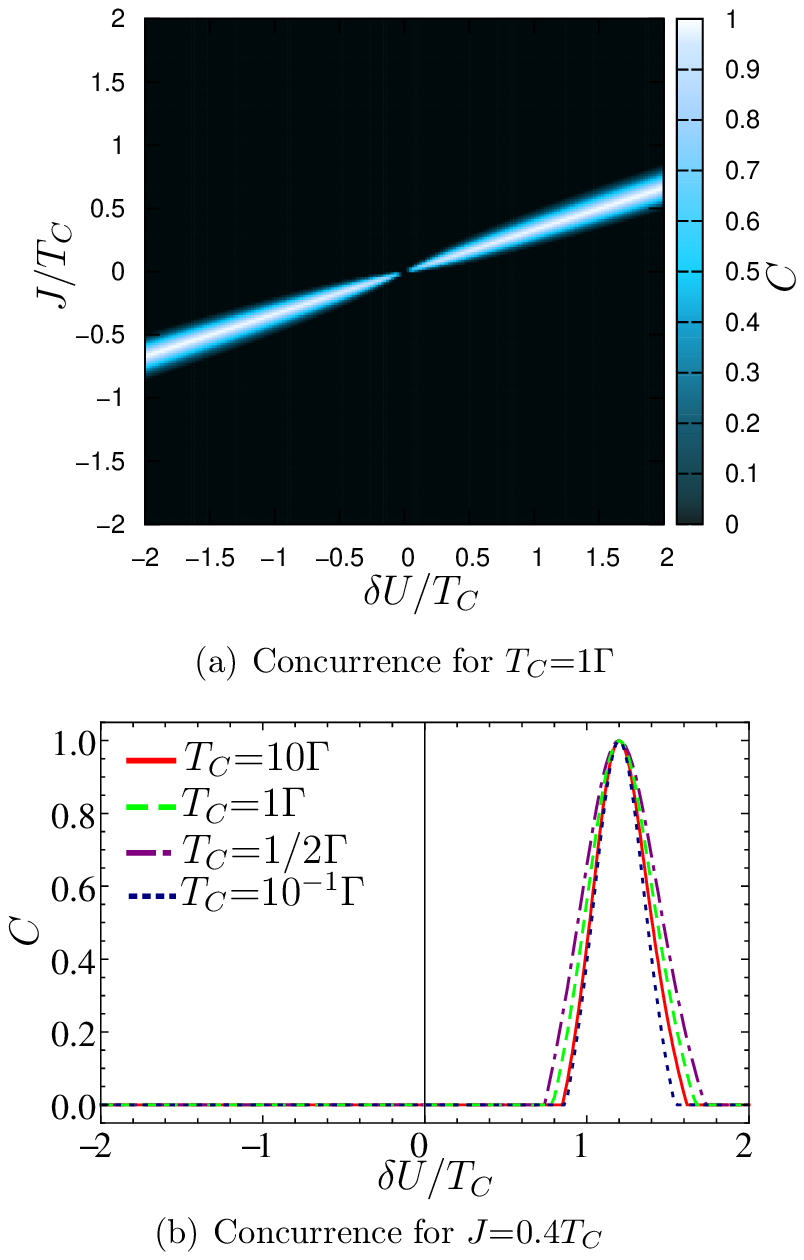}
\caption{(a) Concurrence as function of the exchange energy $J$  and $\delta U$ normalized by $T_C$ for $T_C=1\Gamma$. (b) Concurrence as a function of $\delta U$ normalized by $T_C$ at $J=0.4T_C$ for different ratios of  $T_C/\Gamma$. Parameter: $V=10T_C$.   }
\label{fig:concur}
\end{figure} 

\section{Concurrence \label{sec:concur}} 
Entanglement is a very important aspect of quantum mechanics. 
It is responsible for the non-locality of quantum mechanics, which can be tested\cite{Bell-in-te,Bell-in-exp} 
via the violation of the Bell's inequality.\cite{Bell-in} Apart    
from that,
coherences are the foundation of various concepts in quantum mechanics such as quantum computation,\cite{DiVin-qu-Com,nielsenchuang}  quantum teleportation \cite{BBC93-Tele,BJM97-tele-exp,nielsenchuang} and quantum cryptography.\cite{Eke91-qu-cry,nielsenchuang}

A way to measure the entanglement of a mixed state is to calculate its {\it concurrence}\cite{wootters98} $C$. The concurrence of a two-qubit system is define as $C=\max[0,\sqrt{\lambda_1}-\sum_{j=2}^4\sqrt{ \lambda_j}]$, with $\lambda_j$ being the eigenvalues of 
$\rho_{\text{2Q}}(\sigma_y \otimes \sigma_y)\rho_{\text{2Q}}^* (\sigma_y \otimes \sigma_y) $ in decreasing order. Here,  $\rho_{\text{2Q}}$ is the density matrix of the two qubit system in the localized basis and $\sigma_y$ are the corresponding Pauli matrices of the qubits. The spin degree of freedom is entangled  in the two-TQD setup. In order to calculate the concurrence with respect to the spin qubits, it is necessary to trace out the QD states of the stationary state $\rho_{\text{stat}}$ of \eq{TQD-ent:Trans_Lin} $\rho_{\text{spin}}=\text{Tr}_{\text{QD}}[\rho_{\text{stat}}] $. The two-particle sector of $\rho_{\text{spin}}$ corresponds then to $\rho_{\text{2Q}}$.\cite{foot3}

Fig.~\ref{fig:concur}(a) shows the concurrence as density plot for the same parameters as \ref{Fig:2-TQD-DS-CF}(a). At the pure entangled singlet-DS, the concurrence rises to one, indicating a maximal entangled states. Around the mixed triplet-DS, the concurrence is zero, which corresponds to a state without entanglement. 
\fig{fig:concur}(b) shows the concurrence as a function of $\delta U /T_C$ at $J=0.4T_C$ for different ratios of $T_C/\Gamma$. As for the corresponding Fano factor, shown in \fig{Fig:2-TQD-DS-CF}(d), the concurrence resonance is widest for $T_C=\textstyle{\frac{1}{2}} \Gamma$.

\section{Conclusions}

We have shown that the preparation of a spin-entangled two-electron DS is possible. To be more precise, the system runs simply into a spin-entangled DS as the steady state once the conditions for the spin-entangled DS formation are fulfilled. With this aim, we have introduced a setup with two TQDs. The isotropic spin-exchange interaction between the occupying electrons lifts the degeneracy of the singlet-DS with the three triplet-DSs and enables the creation of a pure spin-entangled singlet-DS. The concurrence, which rises to unity at the singlet-DS, proves the existence of the entanglement in the transport regime.   
Furthermore, the singlet-DS has a strongly enhanced Fano factor compared to the triplet-DS. This signature of the singlet-DS can be used to separate the entangled DS from the non-entangled-DS by measuring the Fano factor.

As the electrons are still localized in the TQDs, this setup enables the creation of spatially separated spin-entangled electrons on demand without any further modifications of the device the entangled electrons are simply stored in the two-TQD setup. 
Switching the chemical potential of the sources leads, such that they also become collectors, enables the usage of the entangled electrons outside of the device. Here, the disadvantage is that each electron has two possibilities to tunnel out of the device, namely, the two former source leads of the TQD it is occupying.
Theoretically, this can be easily avoided by switching the tunnel rate of one of the former sources of each TQD to zero, e.g., $\Gamma_{2a}=\Gamma_{2b}=0$.
Experimentally, it would probably be easier to consider a setup which has only one source and one drain lead for both bias configurations.  
This does not change the essential features as the DS formation and the values of the Fano factor and concurrence around the DSs.

\section*{Acknowledgment}
We are grateful to G.~Platero and F.~Renzoni for helpful discussions. Financial support by DFG Projects GRK 1558, DFG BR 1528/7-1, DFG BR 1528/8-1 and SFB 910 is acknowledged.

\appendix  
\begin{widetext}
\section{Hamiltonian}\label{App:2TQD-DS}
The system Hamiltonian of the two {TQD}s above each other has a block structure
 \begin{align}
\hat{H}_D= 
\begin{pmatrix}
                         0 & 0 & 0 & 0 & 0 & 0 & 0 & 0 & 0\\
0 & H_{\uparrow a}           & 0 & 0 & 0 & 0 & 0 & 0 & 0 \\
0 & 0 & H_{\downarrow a}         & 0 & 0 & 0 & 0 & 0 & 0 \\
0 & 0 & 0 & H_{\uparrow b}           & 0 & 0 & 0 & 0 & 0 \\
0 & 0 & 0 & 0 & H_{\downarrow b}         & 0 & 0 & 0 & 0 \\
0 & 0 & 0 & 0 & 0 & H_{\uparrow a \downarrow b}-\bar{J} & 2\bar{J}      & 0 & 0 \\
0 & 0 & 0 & 0 & 0 & 2 \bar{J} & H_{\downarrow a \uparrow b }-\bar{J}     & 0 & 0 \\
0 & 0 & 0 & 0 & 0 & 0 & 0 &H_{\uparrow a \uparrow b}+\bar{J}       & 0 \\
0 & 0 & 0 & 0 & 0 & 0 & 0 & 0 &H_{\downarrow a \downarrow b}+\bar{J} 
\end{pmatrix},
\end{align}
where the zero in the first diagonal entry denotes the empty state, and the next four entries of the form $H_{\sigma A}$ denote the single-particle sectors of the two-{TQD} system, with  $A\in\{a,b\} $ labeling the {TQD} and $ \sigma\in\{\uparrow, \downarrow\}$  labeling the spin of the electron. In the basis $\{\ket{1_{A\sigma}},\ket{2_{A\sigma}},\ket{3_{A\sigma}}\}$, these part have the form
   \begin{align}
H_{\sigma A}= 
\begin{pmatrix}
 \Delta_A & 0 & T_A \\ 
 0 & - \Delta_A  & T_A \\
T_A & T_A & 0 
\end{pmatrix}.
\end{align}
Here, $T_A$ is the coupling term between {QD} $1$ and QD $3$, and QD $2$ and QD $3$, and $2 \Delta_A$ a detuning between the first and the second dot, $E_{1,A}=\Delta_A $, $E_{2,A}=-\Delta_A $ and $E_{3,A}=0$.   
The last  four diagonal terms $H_{\sigma a \sigma' b}$ are two-particle sectors. In the basis $$\{\ket{1_{a\sigma} 1_{b\sigma'}},\ket{1_{a\sigma} 2_{b\sigma'}},\ket{1_{a\sigma} 3_{b\sigma'}},\ket{2_{a\sigma} 2_{b\sigma'}},\ket{2_{a\sigma} 1_{b\sigma'}},\ket{2_{a\sigma} 3_{b\sigma'}},\ket{3_{a\sigma} 3_{b\sigma'}},\ket{3_{a\sigma} 1_{b\sigma'}},\ket{3_{a\sigma} 2_{b\sigma'}}\},$$  
the Hamiltonians become 
   \begin{align}
H_{\sigma a \sigma' b}= \left(
 \begin{smallmatrix}
 U_{11}+\Delta _a+\Delta _b & 0 & T_b & 0 & 0 & 0 & 0 & T_a & 0 \\
 0 & U_{12}+\Delta _a-\Delta _b & T_b & 0 & 0 & 0 & 0 & 0 & T_a \\
 T_b & T_b & U_{13}+\Delta _a & 0 & 0 & 0 & T_a & 0 & 0 \\
 0 & 0 & 0 & U_{22}-\Delta _a-\Delta _b & 0 & T_b & 0 & 0 & T_a \\
 0 & 0 & 0 & 0 & U_{21}-\Delta _a+\Delta _b & T_b & 0 & T_a & 0 \\
 0 & 0 & 0 & T_b & T_b & U_{23}-\Delta _a & T_a & 0 & 0 \\
 0 & 0 & T_a & 0 & 0 & T_a & U_{33} & T_b & T_b \\
 T_a & 0 & 0 & 0 & T_a & 0 & T_b & U_{31}+\Delta _b & 0 \\
 0 & T_a & 0 & T_a & 0 & 0 & T_b & 0 & U_{32}-\Delta _b
 \end{smallmatrix} \right)
.
\end{align}
The $\bar{J}$ denotes isotropic exchange energy terms, with
   \begin{align}
\bar{J}= 
 \begin{pmatrix}
 J_1 & 0&0& 0&0&0& 0&0&0 \\
0 &  J_2&0& 0&0&0& 0&0&0 \\
0 & 0& J_2& 0&0&0& 0&0&0 \\
0 & 0&0&  J_1&0&0& 0&0&0 \\
0 & 0&0& 0& J_2&0& 0&0&0 \\
0 & 0&0& 0&0& J_2& 0&0&0 \\
0 & 0&0& 0&0&0&  J_1&0&0 \\
0 & 0&0& 0&0&0& 0& J_2&0 \\
0 & 0&0& 0&0&0& 0&0& J_2 \\
 \end{pmatrix}
.
\end{align}
The off-diagonal $J$ terms in the Block structure Hamiltonian switch the spin of the electrons and the diagonal terms change the effective charging energy of the sector.  

\section{Coupling terms}\label{App:2TQD-DS-B}
We find 12 coupling terms $D_X$ in the Lindblad equation \eq{TQD-ent:Trans_Lin} of the two-TQD setup: 
\begin{align}
D_{3 a\sigma} &= \sqrt{\Gamma_{3 a\sigma}}\big(\op{0}{3_{a\sigma}}+\sum_{\sigma'} \big(\op{1_{b \sigma'}}{3_{a \sigma }1_{b \sigma'}}+\op{2_{b \sigma'}}{3_{a\sigma}2_{b \sigma'} }+\op{3_{b \sigma'}}{3_{a\sigma} 3_{b \sigma'}}\big), \nonumber
\\
D_{3 b\sigma} &= \sqrt{\Gamma_{3 b\sigma}}\big(\op{0}{3_{b\sigma}}-\sum_{\sigma'} \big(\op{1_{a \sigma'}}{1_{a \sigma'}3_{b \sigma }}+\op{2_{a \sigma'}}{2_{a \sigma'} 3_{b\sigma}}+\op{3_{a \sigma'}}{3_{a \sigma'} 3_{b\sigma}}\big), \nonumber
\\
D_{1 a\sigma}^\dagger &=\sqrt{\Gamma_{1 a\sigma}}\big(\op{0}{1_{a\sigma}}+\sum_{\sigma'} \big(\op{1_{b \sigma'}}{1_{a \sigma }1_{b \sigma'}}+\op{2_{b \sigma'}}{1_{a\sigma}2_{b \sigma'} }+\op{3_{b \sigma'}}{1_{a\sigma} 3_{b \sigma'}}\big), \nonumber
\\
D_{1 b\sigma}^\dagger &=\sqrt{\Gamma_{1b\sigma}}\big(\op{0}{1_{b\sigma}}-\sum_{\sigma'} \big(\op{1_{a \sigma'}}{1_{a \sigma'}1_{b \sigma }}+\op{2_{a \sigma'}}{2_{a \sigma'} 1_{b\sigma}}+\op{3_{a \sigma'}}{3_{a \sigma'} 1_{b\sigma}}\big), \nonumber
\\
D_{2 a\sigma}^\dagger &=\sqrt{\Gamma_{2a\sigma}}\big(\op{0}{2_{a\sigma}}+\sum_{\sigma'} \big(\op{1_{b \sigma'}}{2_{a \sigma }1_{b \sigma'}}+\op{2_{b \sigma'}}{2_{a\sigma}2_{b \sigma'} }+\op{3_{b \sigma'}}{2_{a\sigma} 3_{b \sigma'}}\big), \nonumber
\\
D_{2 b\sigma}^\dagger &=\sqrt{\Gamma_{2b\sigma}}\big(\op{0}{2_{b\sigma}}-\sum_{\sigma'} \big(\op{1_{a \sigma'}}{1_{a \sigma'}2_{b \sigma }}+\op{2_{a \sigma'}}{2_{a \sigma'} 2_{b\sigma}}+\op{3_{a \sigma'}}{3_{a \sigma'} 2_{b\sigma}}\big).
\end{align}
In the following, we set $\Gamma_{1A\sigma}=\Gamma_{1A}$, $ \Gamma_{2A\sigma}=\Gamma_{2A}$, and $\Gamma_{3A\sigma}=\Gamma_{3A}$, $A=a,b$, $\sigma=\uparrow,\downarrow$. We have assumed energy-independent rates $\Gamma_{iA\sigma}=2\pi \sum_k| V_{\alpha k \sigma,A}|^2 \delta(\omega-\varepsilon_{\alpha k,A})$. 
\end{widetext}

%

\begin{thebibliography}{10}
\bibitem{foot1}{A 
more elaborate model for an exchange interaction is the Hubbard approximation (e.g. Ref.~[\onlinecite{HKP07-spin,LD-Com-QDs}]) on the spin dynamics, with a finite tunnel amplitude $T_{ab}$ between TQD $a$ and TQD $b$. The strength of the exchange interaction is $J_{ij}\sim T_{ab}^2/U_{AA}$, with $U_{AA}$ being the charging energy for finding two electrons in one TQD. In our setup where we assume $T_{ab}\rightarrow 0$ and $U_{AA}\gg U_{i j }$, this would correspond to small exchange interactions. Note that a finite coupling strength $T_{ab}$ would not affected the entanglement of the two electrons since both TQDs are in the strong Coulomb blockade regime. Hence only a single excess electron is allowed in each TQD.  However, a finite $T_{ab}$ would lead to a finite probability for the two entangled electrons to tunnel out of the same TQD, as it would enable the tunneling of an electron from TQD $a$ to TQD $b$ once TQD $b$ is unoccupied, or vice versa. Therefore, the perfect separation of the entangled electrons would be lost. This 
can be avoided by modifying the strength of $J_{ij}$ depending on 
the occupation of the system, e.g. via the applied gate voltages.\cite{LD-Com-QDs}    
}
\bibitem{foot-M}{The set of equations given by \eq{DS-condition-1} together with the normalization condition of the DS can be solve by symbolic computational software program (e.g. Mathematica). However, these softwares may miss solutions and also find unphysical or uninteresting (e.g. with the condition $T_a=T_b=0$) solutions. Such that each solution must be checked for its uniqueness and it might be possible that further DSs exist that are not found. }
\bibitem{foot3}{The original definition of the concurrence in Ref.~[\onlinecite{wootters98}] considers only the case where $\text{Tr}[\rho_{\text{2Q}}]=1$. 
As we not only have a two particle sector of $\rho_{\text{spin}}$, but also an empty and a single particle sector, the trace of $\text{Tr}[\rho_{\text{2Q}}]<1$ is, in general, smaller than one. 
Since the 'concurrence' for matrices with a different trace scales linearly with the value of the trace, we can also calculate $C$ for a matrix with $\text{Tr}[\rho_{\text{2Q}}]<1$ to measure the degree of entanglement. 
In the limit of large left tunnel rates and when the system is in the dark state, the trace becomes unity anyway.}
\end{thebibliography}

\end{document}